\documentclass{article}
\usepackage{arxiv}
\usepackage[utf8]{inputenc} 
\usepackage[T1]{fontenc}  
\usepackage{hyperref}       
\usepackage{url}            
\usepackage{booktabs}       
\usepackage{nicefrac}       
\usepackage{microtype}      
\usepackage{lipsum}
\usepackage{cite}
\usepackage{caption}
\usepackage{amsmath,amssymb,amsfonts}
\usepackage{algorithmic}
\usepackage{graphicx}
\usepackage{textcomp}
\usepackage{graphics} 
\usepackage{epsfig} 
\usepackage{mathptmx} 
\usepackage{times} 
\usepackage{amsmath} 
\usepackage{amssymb}  
\usepackage{lscape}
\usepackage{multirow}
\usepackage{bm}
\usepackage{url}
\usepackage{diagbox}
\usepackage[ruled,vlined,linesnumbered]{algorithm2e}
\usepackage{soul}
\usepackage{color}

\title{Analysis of MRI Biomarkers for Brain Cancer Survival Prediction}

\author{
    Subhashis~Banerjee\\
    Machine Intelligence Unit\\
    Indian Statistical Institute\\
    Kolkata 700108, India\\
    \texttt{mail.sb88@gmail.com}\\
    
    \And
    
    Sushmita~Mitra\\
    Machine Intelligence Unit\\
    Indian Statistical Institute\\
    Kolkata 700108, India\\
    \texttt{sushmita@isical.ac.in}\\
    
    \And
    
    Lawrence~O.~Hall\\
    Department of Computer Science and Engineering\\
    University of South Florida\\
    4202 E. Fowler Ave., Tampa, FL. 33620-9951, United States\\
    \texttt{lohall@mail.usf.edu}\\
}

\begin{document}
\maketitle
\begin{abstract}
	Prediction of Overall Survival (OS) of brain cancer patients from multi-modal MRI is a challenging field of research. Most of the existing literature on survival prediction is based on Radiomic features, which does not consider either non-biological factors or the functional neurological status of the patient(s). Besides, the selection of an appropriate cut-off for survival and the presence of censored data create further problems. Application of deep learning models for OS prediction is also limited due to the lack of large annotated publicly available datasets. In this scenario we analyse the potential of two novel neuroimaging feature families, extracted from brain parcellation atlases and spatial habitats, along with classical radiomic and geometric features; to study their combined predictive power for analysing overall survival. A cross validation strategy with grid search is proposed to simultaneously select and evaluate the most predictive feature subset based on its predictive power. A Cox Proportional Hazard (CoxPH) model is employed for univariate feature selection, followed by the prediction of patient-specific survival functions by three multivariate parsimonious models \textit{viz.} Coxnet, Random survival forests (RSF) and  Survival SVM (SSVM). The brain cancer MRI data used for this research was taken from two open-access collections TCGA-GBM and TCGA-LGG available from The Cancer Imaging Archive (TCIA). Corresponding survival data for each patient was downloaded from The Cancer Genome Atlas (TCGA). A high cross validation $C-index$ score of $0.82\pm.10$ was achieved using RSF with the best $24$ selected features. Age was found to be the most important biological predictor. There were $9$, $6$, $6$ and $2$ features selected from the parcellation, habitat, radiomic and region-based feature groups respectively. Analysis of each selected feature was performed individually using Kaplan-Meier survival curves, pair-wise Spearman rank correlation and heat maps.
\end{abstract}


\section{Introduction} \label{sec:introduction}
Gliomas are the most common primary brain tumors originating from the glial cells in the Central Nervous System. Based on histology gliomas are broadly classified into two categories, viz. Low-grade gliomas (LGG) and high-grade gliomas (HGG) or glioblastoma (GBM) \cite{louis16,Louis2021}. The HGG encompasses grades III and IV of the WHO categorization, and exhibits a rapid proliferating behaviour with a short overall survival (OS)\footnote{The average length of survival time from either the date of diagnosis or the start of treatment for a disease} time of only $14$ months. Diffuse LGG are infiltrative brain neoplasms which include the histological classes astrocytomas, oligodendrogliomas, and oligoastrocytomas. Although in LGG patients longer overall survival is expected as compared to HGG,  some molecular subtypes of LGG result in an outcome similar to glioblastomas (WHO grade IV) and eventually progress to secondary GBMs \cite{liyun13}. 

Recently five new molecular subtypes of gliomas have been announced by the World Health Organization (WHO) \cite{louis16} based on histologic phenotypes as well as the genotypes of the tumor. Mutation in the ``isocitrate dehydrogenase'' (IDH) gene and the combined loss of the short arm of chromosome 1 (i.e. 1p) and the long arm of chromosome 19 (i.e. 19q) ``1p/19q codeletion'' is recognized as a crucial genetic factor along with the histological grading. Among the five molecular subtypes, the foremost favorable clinical outcome and presumably improved sensitivity to treatment is observed in the IDH mutated and 1p/19q codeleted glioma subtype \cite{cairncross2013phase}. It is found to be related to longer survival, particularly for oligodendrogliomas which are more sensitive to chemotherapy. On the other hand, IDH wild type gliomas have an outcome similar to the glioblastomas (WHO grade IV). Intermediate prognostic outcomes (far better than LGG-wild) are observed in IDH mutated and 1p/19q non-codeleted gliomas. IDH mutated HGG are considered to be secondary GBM, whereas IDH wild type HGG correspond to primary GBM. 

Magnetic Resonance Imaging (MRI) has become the standard non-invasive technique for brain tumor diagnosis over the last few decades, due to its good soft tissue contrast \cite{8674823}. MRI has the advantage of being able to scan the entire tumor in vivo and can demonstrate a strong correlation with histological grade. Multi-sequence MRI plays a major role in the detection, diagnosis, and management of brain cancers in a non-invasive manner. Recent literature reports that computerized detection and diagnosis of the disease, based on medical image analysis, could be a good alternative or addition to the manual mode of radiologists. MR imaging can effectively capture the intrinsic heterogeneity of gliomas using multimodal scans with varying intensity profiles. Typically four MR sequences viz. native $T1$-weighted ($T1$), $T2$-weighted ($T2$), post-contrast enhanced $T1$-weighted ($T1C$) and $T2$-weighted with FLuid-Attenuated Inversion Recovery ($FLAIR$), are used \cite{9122459}.

Decoding of tumor phenotype\footnote{Observable characteristics corresponding to a genotype.} using noninvasive imaging is a recent field of research, known as {\it Radiomics} \cite{rizzo2018radiomics, gillies2015radiomics, 9146338}, and involves the extraction of a large number of quantitative imaging features that may not be apparent to the human eye. An integral part of the procedure involves manual or automated delineation of the 2D region of interest (ROI) or 3D volume of interest (VOI) \cite{9174647, banerjee2018automated, 8287819}, to focus attention on the malignant growth. This is typically followed by the extraction of suitable sets of quantitative imaging features from the ROI or VOI, to be subsequently analyzed through machine learning towards making decisions. Feature selection enables the elimination of redundant and/or less important subset(s) of features, for improvement in speed and accuracy. This is particularly relevant for high-dimensional radiomic features, extracted from medical images. 

Recently prediction of Overall Survival (OS) of patients, from multi-modal MRI, has gained a lot of attention \cite{brats2018}. The majority of the models proposed in the literature for survival prediction are based on radiomics \cite{feng2018brain, puybareau2018segmentation, sun2018tumor, weninger2018segmentation, banerjee2018multi}. Application of deep learning models for OS prediction is limited. 
As reported in Ref. \cite{suter2018deep}, a very low testing accuracy ($51.5\%$) was achieved using the best performing deep learning model. In this scenario the limited size of annotated datasets, containing appropriate survival information, makes it infeasible to apply deep learning methods for survival prediction. Therefore the state-of-the-art methods try to predict the OS based on quantitative imaging features only, whether manually extracted or automatically learned. Typically the OS prediction problem is transformed into a classification problem by dividing the patients into survival groups \textit{viz.} long, mid, and short \cite{brats2018}. A major challenge, however, lies in determining appropriate thresholds(s). The cut-off defined over a sample population may not be appropriate for a larger (or smaller) population size. Therefore we consider survival as a regression problem in this research. Moreover, in the presence of censored data it is not desirable to further reduce the size of the dataset. Regression, considering censoring, is called survival regression and falls under survival analysis.         

The problem of predicting OS in brain cancer patients is challenging. It depends on multiple factors like cancer grade, molecular biomarkers (IDH mutation \& 1p/19q codeletion status), non-biological factors (tumor location, whether it was surgically removed or reduced), age, sex and other comorbidities. Although quantitative MRI features have been found to be useful for predicting tumor grade \cite{banerjee2017fuzzieee, banerjee2018brain} or molecular biomarkers \cite{lu2018machine}, they do not consider non-biological factors or the functional neurological status of the patient(s).

In order to circumvent this problem we introduce two novel neuroimaging feature families, extracted from a brain parcellation atlas and spatial habitats, along with the classical radiomic and region-based features; to study their combined predictive power for survival analysis. Radiomic features typically attempt to capture lesion heterogeneity through quantitative imaging descriptors. As such, they can be useful predictors for the tumor grade or for determining different molecular biomarkers like IDH1/2 mutations and/or 1p/19q co-deletion status. Brain parcellation atlas-based features capture the distribution of tumor cells in different regions of the brain, thereby holding a key to evaluating prognosis. The spatial heterogeneity within a tumor is an important indicator of survival and prognosis, and can be modeled using tumor habitats. Features extracted from tumor habitats allow representation of different biological properties, reflecting heterogeneous tumor subpopulations within the same class. 

The rest of the paper is organized as follows. In Section \ref{section2} we present the details of the proposed neuroimaging features. Section \ref{section3} describes in detail the MRI dataset, its pre-processing, segmentation of volume of interest, extraction of neuroimaging features, and employment of machine learning for predicting overall survival time. Section \ref{section4} reports the experimental setup and validation of survival prediction, along with clinical significance of the features. Finally Section \ref{section5} draws conclusions, and provides directions for future research.

\section{Extraction of Features for Predicting Survival  \label{section2}}
In this section we briefly introduce the four categories of extracted features used for survival analysis of brain cancer patients. These include radiomic features from existing literature, along with the novel region-based and neuroimaging features introduced here.

\subsection{Radiomic features}
Decoding the tumor phenotype using noninvasive imaging, involving evaluation of a large number of quantitative imaging features which may not be apparent to the human eye, is called \emph{Radiomics} \cite{gillies2015radiomics}.  An integral part of the procedure involves manual or automated delineation of the 2D region of interest (ROI) or 3D volume of interest (VOI) \cite{banerjee2016single, banerjee2018automated}, to focus attention on malignant growth. This is typically followed by the extraction of suitable sets of hand-crafted quantitative imaging features from the ROI or VOI, to be subsequently analyzed through machine learning towards decision-making. Different statistics can be computed from the distribution of voxel intensities in MRI, within the tumor volume, for quantitatively estimating the tumor. These include \cite{banerjee2017fuzzieee}
\begin{itemize}
	\item First-order Gray Level Statistic (GLS) based features: provide information about the gray level distribution of voxel intensities within the segmented VOI.
	\item Gray Level Co-occurrence Matrix (GLCM) based features: provide information about the relative position (or association) of discrete gray level intensities in an image. This provides a different way of describing second-order statistical texture-based features, and can expose certain other characteristics about the spatial distribution of gray levels in an image.
	\item Gray Level Run Length (GLRL) matrix based features: quantify the gray level runs in an image. A gray level run is the length, in number of consecutive pixels along a given direction, for a particular gray level value.
	\item Gray Level Size Zone (GLSZ) matrix based features: define gray level zones within the ROI/VOI. The number of 26-connected voxels having the same gray level intensity is considered as a gray level zone.
	\item Geometrical Shape and Size (GSS) based features: provide insight about the shape and size of the tumor VOI. Geometric features refer to the visual interpretation based on the tumor boundary points, and help in its recognition based on shape. 
\end{itemize}

Here the five categories of radiomic features  refer to the textural properties of an image, with GLCM, GLRL and GLSZ constituting the structured approach. Texture is characterized by its coarseness and directionality. While the statistical approach describes a quantitative measure of the arrangement of intensities in the VOI, the structured one deals with the repetition of pixel intensities and their relationship. The textural features are described as a function of the spatial distribution of the pixel intensity values in gray scale. The VOI is said to have a constant texture when a set of local statistics or other local properties of the image are either constant, slowly varying, or approximately periodic. The GSS features represent properties like translational-, rotational-, and scale-invariance of an object.

\subsection{Region-based features}
We additionally designed tumor region features, encompassing important details related to prognosis. The tumor region features provide quantitative measurements like volume ratio (of different tumor sub-regions with respect to the entire brain), along with the mean and standard deviation of intensities (of voxels) from the different tumor sub-regions \textit{viz.} peritumoral edema ($ED$), necrotic core ($NCR$), enhancing and non-enhancing tumor core ($ET / NET$), along with several aggregations like tumor core ($TC: NCR/NET + ET$) and whole tumor ($WT:  NCR/NET + ET + ED$). These sub-regions, both individually and collectively, reflect clinically relevant information in terms of the corresponding mean ($\mu$) and standard deviation ($\sigma$). 

The various radiomic and region-based features, used in this study, are summarized in Table S1 of the Supplementary Information section. In the following two subsections we describe the neuroimaging features proposed in this study.

\subsection{Brain parcellation atlas features}\label{sec:brain_parcellation}
In the spatial domain a brain is represented in terms of parcellations \cite{eickhoff2018imaging}, which correspond to non-overlapping regions exhibiting similar properties like Cytoarchitecture (microscopic study of the cellular composition of tissues of the Central Nervous System), anatomical or functional connectivity. A tumor may develop in any part of the brain, with some locations being more damaging (or sensitive) than others and perhaps more difficult to treat. Therefore the distribution of the tumor cells in different brain regions is likely to provide useful prognostic information.  

\begin{figure}[]
	\centering\includegraphics[width=1.0\linewidth]{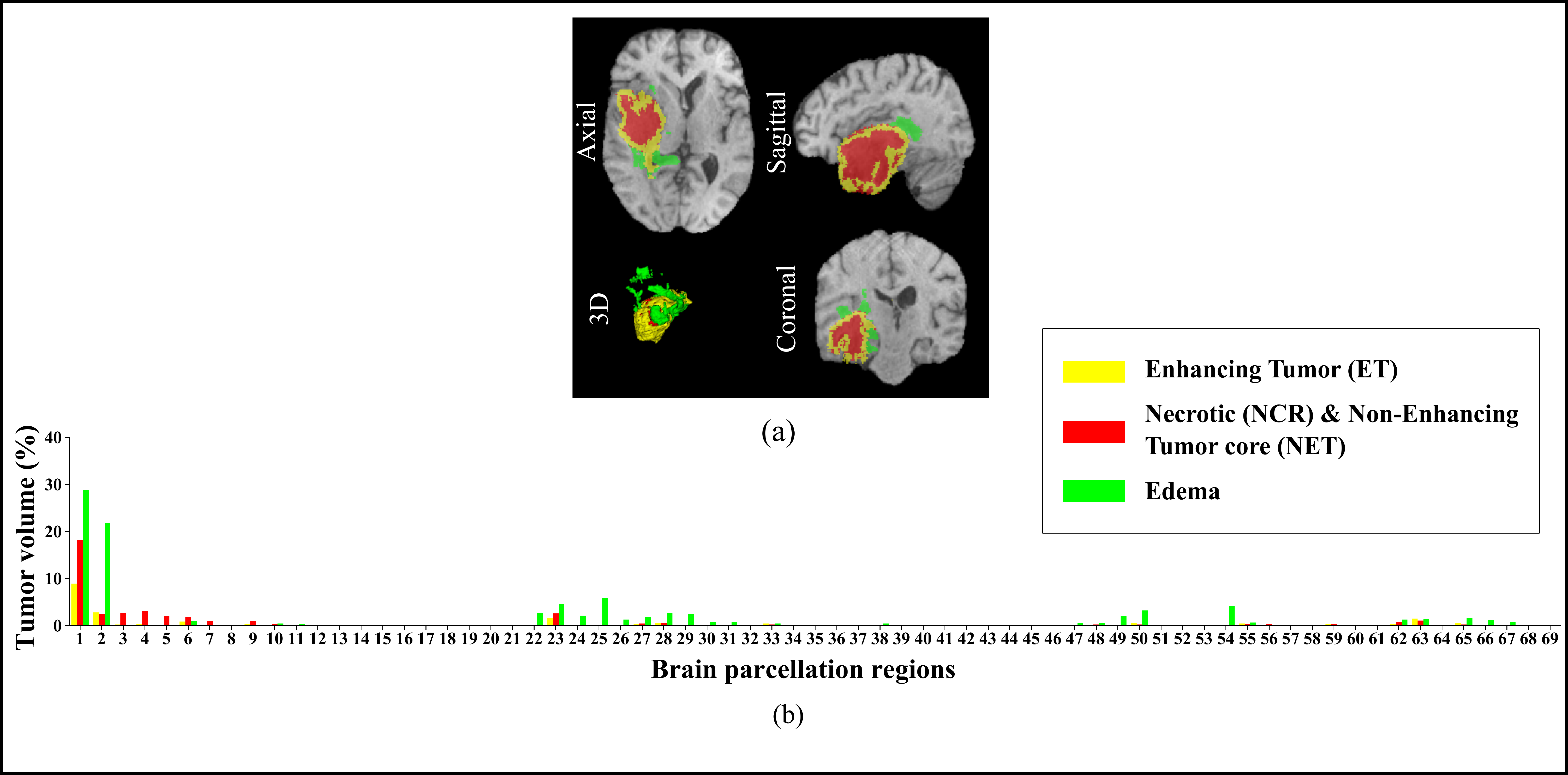}
	\caption{(a) $T2$ MRI scan for a sample patient, with (b) the corresponding tumor cell distribution across the parcellation regions.}
	\label{fig:parcellation_diagram}
\end{figure}

Brain parcellations are typically generated by specific clustering techniques using information extracted over high quality T1-weighted images \cite{desikan2006automated}. The Harvard-Oxford cortical/subcortical atlases, consisting of $48$ cortical and $21$ subcortical structural regions, is a probabilistic atlas generated from $21$ healthy male and $16$ healthy female subjects. The names of the $69$ parcellation regions are provided in Table S2 of the Supplementary Information section. Tumor cell distribution in the MRI volume of a sample patient [shown in Fig. \ref{fig:parcellation_diagram}(a)] is illustrated in Fig. \ref{fig:parcellation_diagram}(b). The three different tumor sub-regions (\textit{viz.} $ET / NET$, $NCR$, and $ED$) volume percentages are computed for each of the $69$ parcellation regions, dividing by the corresponding whole tumor volume. This results in a total of $207$ parcellation features. 

\subsection{Spatial habitat features \label{habitat}}
It is often observed that (epi)genetic properties of individual cancer cells are highly variable, even within the same tumor, such that preexisting resistant clones emerge and proliferate after therapeutic selection targeting the sensitive clones. Such heterogeneity within tumors is found to offer resistance to conventional therapies, by accelerating the unopposed proliferation of resistant subpopulations at the expense of the elimination of corresponding dose-susceptible subpopulations as well as their attendant competition for space and substrate.

Tumors can therefore be modeled as spatially heterogeneous complex adaptive systems \cite{dextraze2017spatial}, in which tumor growth and response to therapy are governed by eco-evolutionary interactions between the tumor micro-environment and phenotypic properties of local cellular subpopulations. Temporal and spatial cellular heterogeneities are due to clonal evolution from accumulating random mutations in cancer cell populations. Volumetric differential imaging holds promise in characterizing tumor heterogeneity at a holistic level, as compared to tissue biopsy which is constrained to sampling only a small fraction of the tumor.

Spatial heterogeneity within a tumor occurs mainly due to variations in cell density, necrosis, blood flow, etc. Such collection of locally homogeneous tumor regions constitute the habitats; and can be formulated in terms of their varying intensity profiles when using multimodal MRI scans (like $T1$, $T1C$, $T2$, $FLAIR$). Each tumor is thereby quantified as some combination of distinct habitats. Therefore habitat imaging, combining radiomics with multiparametric imaging, has the potential to provide noninvasive longitudinal biomarkers of intratumoral evolutionary and ecological dynamics for the informed application of adaptive therapy to manage tumors.

\begin{figure}[]
	\centering\includegraphics[width=1.0\linewidth]{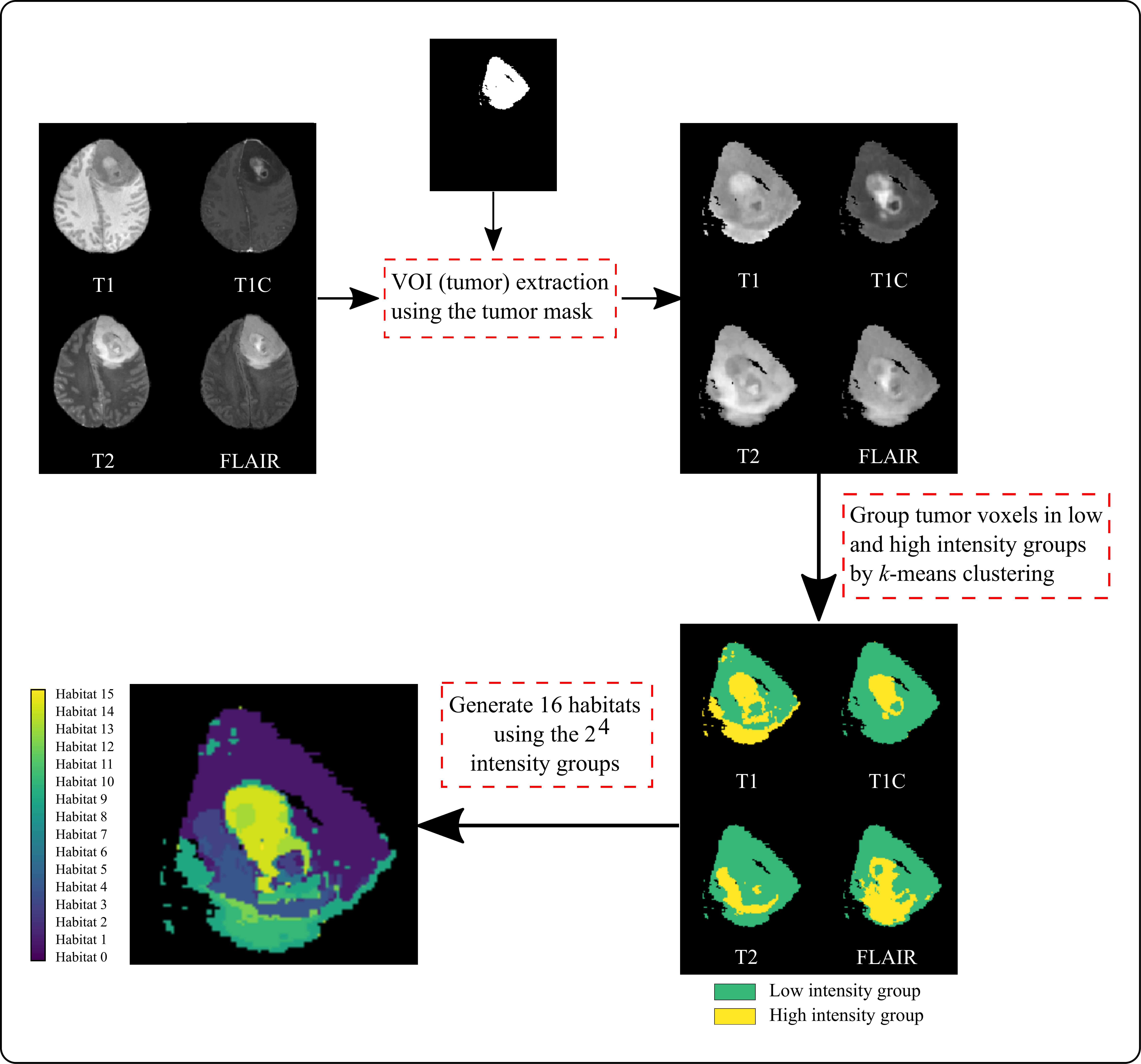}
	\caption{Generating $16$ spatial habitats, based on the four MR sequences.}
	\label{fig:habtat_generation}
\end{figure}

In this research the tumor voxels were grouped into high and low intensity groups, based on the voxel intensity, by employing $k$-means clustering over each MR sequence. A total of $16$ imaging habitats were generated from the four MR sequences, by considering all possible combinations of high and low intensity groups. The $16$ habitats are represented by a string of four binary numbers from 0000 to 1111, where ``0'' and ``1'' represent the low and high intensity groups in each of the four MRI sequences in this order ``$T1-T1C-T2-FLAIR$''. Next this binary representation was converted to decimal form, to be used as the label of the habitat. Thereby habitat $H_0$ represents the region where all four MR sequences have low intensity (0000), and $H_{15}$ corresponds to the region where all the MR sequences have high intensity (1111). Fig. \ref{fig:habtat_generation} demonstrates the steps to obtain the $16$ habitats, within the tumor region, from the four MR sequences. Thereby $16$ volumetric habitat features, quantifying the volumes of the $16$ habitats, are extracted.

\section{Materials and Methods \label{section3}}
In this section we provide a detailed description of the MRI dataset used for our experiments, along with the preprocessing. Segmentation of different tumor sub-regions, extraction of novel neuroimaging features from them, pre-processing of the features, and their selection towards survival analysis are discussed.

\subsection{Dataset preparation}
Brain cancer MRI data used for this research was taken from two open-access collections TCGA-GBM \cite{scarpace9radiology} and TCGA-LGG \cite{pedano2016radiology}, available from The Cancer Imaging Archive (TCIA) \cite{clark2013cancer}. Corresponding survival data for each patient was downloaded from The Cancer Genome Atlas (TCGA) \cite{tomczak2015cancer}. Survival data is represented by a 3-tuple $(X, T, \delta)$, where $X$ is an $n\times p$ matrix containing $p$ neuroimaging features (extracted from the $n$ samples considered in the study), $T$ is an $n$ dimensional vector with the observed survival (or censored) time for each patient, and $\delta \in\{0, 1\}$ represents the corresponding $n$ censoring\footnote{Censoring is a condition in which the value of a measurement or observation (say, survival) is only partially known.} indicators.

While the TCGA-GBM database encompasses MR images from $262$ HGG patients, the TCGA-LGG database contains MR images from $199$ LGG patients. For each sample patient the inclusion criteria was the availability of (i) preoperative scans for all the four modalities ($T1$, $T1C$, $T2$, and $FLAIR$), and (ii) corresponding clinical information, including overall survival time and age at diagnosis. Thereby a total of $243$ ($135$ GBM and $108$ LGG) subjects were included in the experiments. The study population demographics are reported in Table \ref{tab:1}.

\begin{table}[]
	\centering
	\caption{Study population demographics from TCGA database.}
	\begin{tabular}{lll}
		\hline
		Gender                      & Male              & 136                            \\
		& Female            & 107                            \\
		Histological type           & Glioblastoma      & 135                            \\
		& Oligodendroglioma & 48                             \\
		& Astrocytoma       & 32                             \\
		& OligoAstrocytoma  & 28                             \\
		Neoplasm histologic grade   & IV                & 135                            \\
		& III               & 59                             \\
		& II                & 49                             \\
		Age at initial diagnosis   & Min               & 17.5                           \\
		& Max               & 84.8                           \\
		& Mean $\pm$ std      & 52.90 $\pm$ 15.50 \\
		& Median            & 54.8                           \\
		Survival (censored) & $>$ 12 months   & 127 (58)                       \\
		& $<$ 12 months   & 116 (50)                       \\ \hline
	\end{tabular}
	\label{tab:1}
\end{table}

The MRI volumes are first co-registered to the same anatomical template, interpolated to $1mm^3$ voxel resolution and skull-stripped, before being used for experimentation.

\subsection{Segmentation and feature extraction}
The first step in building any machine learning model for OS prediction in brain cancer is the segmentation of the VOI; which includes the glioma and its different sub-regions \textit{viz.} ED, NCR, ET and NET, from multi-modal MR images of the brain. This is typically followed by the extraction of suitable sets of radiomic features from the VOI, for subsequent application of machine learning towards decision-making. 

We used a deep learning model, the Multi-Planar Spatial Convolutional Neural Network (MPS-CNN) \cite{sb_frontiers, banerjee2018multi}, for segmenting the glioma (along with its different sub-regions) from multi-modal MR images of the brain. The MPS-CNN was ranked among the top performing methods in the BraTS 2018 challenge doing very well on unseen images. It is an encoder-decoder type ConvNet model, which performs pixel-wise segmentation of the tumor along three anatomical planes (axial, sagittal and coronal) at the slice level. These are then combined, using a consensus fusion strategy with a fully connected Conditional Random Field (CRF) based post-refinement, to remove small isolated regions from the segmented output from the final volumetric segmentation of the tumor and its constituent sub-regions. It is experimentally demonstrated in Ref. \cite{banerjee2018multi} that a significant improvement in the segmentation performance occurs if we include the CRF based post-refinement step. Novel concepts such as spatial-pooling and unpooling were introduced to preserve the spatial locations of the edge pixels, for reducing segmentation error around the boundaries. A new aggregated loss function was also introduced for effectively handling data imbalance \cite{sb_frontiers}.


Radiomic features from the GLS ($19$), GLCM ($24$), GLRL ($16$), GLSZ ($16$) and GSS ($17$) groups were extracted over all four MRI sequences ($T1$, $T1C$, $T2$, $FLAIR$) from the segmented WT volume. This results in the generation of a total of $\mathbf{368}$ radiomic features. Five volume ratios are computed from the region features, with the remaining ten mean and standard deviation based features being extracted over all four MRI sequences. Therefore a total of ($5 + 10 \times 4$) $\mathbf{45}$ tumor Region features, as enumerated in Table S1 of the Supplementary Information section, are generated. The volume fractions of each parcellation region, occupied by the three tumor sub-regions ($ET$, $NCR/NET$, and $ED$), are next computed. We have three such features from each of the $69$ ($48$ cortical, $21$ subcortical) parcellation regions, resulting in a total of $\mathbf{207}$ ($69\times3$) brain parcellation atlas features (as described in Sec. \ref{sec:brain_parcellation}). The $16$ volumetric habitat features, quantifying the volumes of the $16$ habitats, are also computed.

Thus a total of $\mathbf{636}$ ($368+45+207+16$) neuroimaging features were extracted. The components age and sex were included as the biological features. Thereafter, the constant, quasi-constant features are eliminated (leaving 445);  clusters of highly correlated features were collapsed into one representative attribute possessing the largest inter-subject variability or highest dynamic range (leaving 171) \cite{gillies2015radiomics, james2013introduction, voorhees1986implementing}. Adding the two biological features resulted in the final set of $\mathbf{173}$ features.

\subsection{Neuroimaging signature construction}
After preprocessing  the data we are left with non-correlated (potentially) important features for each of the $243$ glioma samples. In order to help avoid overfitting it may be useful to select a minimum subset of features by reducing the feature space according to a predetermined evaluation criterion. Hence each feature was individually evaluated, using the Cox Proportional Hazard (CoxPH) regression model \cite{cox1972regression}, to estimate its impact on survival.

CoxPH is a semi-parametric model which tries to estimate the conditional probability $h_i(t|X_{i,\ast})$ (hazard function) for an individual $i$, such that the individual will experience an event (for example, death) within a small time interval $[t, t+\Delta t)$ given that the patient has survived up to time $t$. The hazard function is represented as
\begin{equation}
h_i(t|X_{i,\ast}) = h_0(t)\exp\Big(\sum_{k=1}^pX_{i, k}W_k\Big),
\end{equation}
where $h_0$ is the baseline hazard function and $W$ is a $p$ dimensional vector of regression coefficients. 

\subsection{Survival analysis}
Next, three multivariate parsimonious models \textit{viz.} (i) Coxnet \cite{simon2011regularization}, (ii) Random survival forests (RSF) \cite{ishwaran2008random}, and (iii) Survival SVM (SSVM) \cite{polsterl2015fast} were
used for survival risk modeling, by excluding the less predictive features obtained from the CoxPH based on the training set.    

Coxnet fits a penalized multivariate CoxPH regression with an elastic-net regularization having both lasso ($\|W_i\|$) and ridge ($\|W_i\|^2_2$) regression penalties. The lasso penalty forces the model to select a few non-zero non-collinear covariates or features, whereas ridge regression uses all the features by scaling less important features towards zero. A balancing factor $0\le\alpha\le1$ is used to adjust the impact of both penalties. The elastic-net regularization can be represented as $\Big(\alpha \sum_{k=1}^p\|W_k\| + \frac{1}{2}(1-\alpha)\sum_{k=1}^{p}\|W_k\|_2^2\Big)$.

Random Survival Forest (RSF) is an ensemble model which combines a set of ``weak'' learners to create a ``strong'' learner. RSFs build upon survival trees by fitting multiple survival trees on subsets of data through a bootstrap approach and averaging results to improve the predictive performance and reduce over-fitting. At each node it selects a random subset of features in order to compute the best features along with a split value $c$ (through the log-rank test) which governs whether each individual sample $i$ goes to the right $(X_i \le c)$ or left $(X_i > c)$ daughter nodes, respectively. Let there be $m$ distinct death times, defined as $t_1 < t_2 < \dots t_m$. The log-rank split-statistic value is defined as

\begin{equation}
L = \frac{\sum_{j=1}^m\Big(d_{j,left} - \beta_{j,left}\frac{d_j}{\beta_j}\Big)}{\sqrt{\sum_{j=1}^{m}\frac{\beta_{j,left}}{\beta_j}\Big(1-\frac{\beta_{j,left}}{\beta_j}\Big)\Big(\frac{\beta_j - d_j}{\beta_j - 1}\Big)d_j}},
\end{equation}
where $d_{j} = d_{j, left} + d_{j, right}$ denotes the number of events at time $t_j$ in left and right daughter nodes, and $\beta_{j} = \beta_{j, left} + \beta_{j, right}$ denotes individuals who are at risk (who have not yet had an event or been censored) at the start of time $t_{j}$ in daughter nodes.

As the separation and classification of patients into high and low risk groups is the primary goal in cancer research, the absolute discrimination power is often not important. Therefore instead of computing the survival probability, it can be modeled as a ranking problem. Here survival analysis is modeled as a classification problem, with the objective of predicting the risk ranks between individuals \cite{van2011support}. The SSVM is employed with the bi-objective function of eqn. (\ref{eqn:ssvm}), optimizing both the regression and ranking objectives. This is expressed as 

\begin{multline}
	\label{eqn:ssvm}
\arg \min_{W, \mathbf{b}} \frac{1}{2} W^T W + \frac{\alpha}{2} \Bigg[
r \sum_{i,j \in \mathcal{P}}
\max(0, 1 - (W^T X_{i,\ast} - W^T X_{j,\ast}))^2\\
+ (1 - r) \sum_{i=0}^n \left( \zeta_{W,\mathbf{b}} (T_i, X_{i,\ast}, \delta_i) \right)^2 \Bigg],
\end{multline}
where 
\begin{equation}
\zeta_{W,\mathbf{b}} (T_i, X_{i,\ast}, \delta_i) =
\begin{cases}
\max(0, T_i - W^T X_{i,\ast} - \mathbf{b}) & \text{if $\delta_i = 0$,} \\
T_i - W^T X_{i,\ast} - \mathbf{b} & \text{if $\delta_i = 1$,} \\
\end{cases}
\end{equation}

\begin{equation}
\mathcal{P} = \{ (i, j)~|~T_i > T_j \land \delta_j = 1 \}_{i,j=1,\dots,n}.
\end{equation}
Here $X_{i,\ast}$ and $X_{j,\ast}$ represent $d$-dimensional feature vectors of two individuals, $T_i$ and $T_j$ denote the survival time or the time of censoring $\delta\in\{0,1\}$. The hyper-parameter  $\alpha>0$  determines the amount of regularization to be applied, while the hyper-parameter $r\in\{0,1\}$  represents the trade-off between the ranking objective and the regression objective. Thus when  $r=1$  it reduces to the ranking objective, and for $r=0$ it boils down to the regression objective.

\section{Experimental Setup and Results \label{section4}}
Since the survival data contains censoring, metrics like root mean squared error or correlation aren't suitable. So we use generalization of the area under the receiver operating characteristic (ROC) curve, called Harrell's concordance index or $C$-index \cite{uno2011c}, which is expressed as

\begin{equation}
C\textrm{-index} = \frac{\sum_{i,j}I(T_i > T_j)I(\hat{Y_i}<\hat{Y_j})\delta_j}{\sum_{i,j}I(T_i > T_j)\delta_j}.
\end{equation}
Here for a pair of subjects $i$, $j$, the survival times are denoted by $T_i$ and $T_j$,
the predicted risk scores are denoted by $\hat{Y_i}$, $\hat{Y_j}$, $\delta_j \in \{0,1\}$ is 0 if the subject is censored, and $I(.)$ is the indicator function. It can be interpreted as the fraction of all pairs of subjects whose predicted risk score are correctly ordered among all subjects that can actually be ordered. If the predicted survival time (probability) is larger for the patient who (actually) lived longer, the prediction for that pair is said to be concordant with the (actual) outcome. Analogous to the $AUC$, a value of $C$-index = 1 corresponds to the best prediction model while $C$-index = 0.5 represents a random prediction.

As the dataset is small and there is no separate testing dataset available, a $k$-fold cross validation ($k=10$) strategy with grid search was used to simultaneously select and evaluate the most predictive feature subset based on its predictive power. The CoxPH model was employed for univariate feature selection over each training fold. This ranked the features according to their $C$-index. The top $p'$ feature subsets (consisting of $1$ to $p'$ number of features) were used to create the three multivariate survival analysis models, viz. Coxnet, SSVM and RSF. These modeled the survival functions while excluding the less predictive features obtained from CoxPH for each training fold. Algorithm~1 outlines the steps. The survival dataset, with reduced number of features ($173$), is represented as $(\mathbf{X'}\in \mathbb{R}^{n\times p'}, \mathbf{T}\in\mathbb{R}^{n}, \mathbf{\delta}\in\{0,1\}^{n})$, where $p' < p$ and $p=636$ and $p'=173$. Finally, the best number of features $\lambda_0$ get selected in Step 12 of the algorithm as the output. The subset of features resulting in the best model is reported in Table \ref{tab:cv}.

The experiments were performed on a laptop with an Intel Core i7 CPU (2.20 Ghz) and 32 GB of RAM. The Feature extraction, pre-processing, and survival analysis codes were developed using Python with open source libraries\footnote{scikit-image, SimpleITK, pandas, scikit-learn, scikit-survival.} The runtime of each of the modules is given in Table \ref{tab:run_time}. For feature extraction we have reported the mean and standard deviation of CPU time for the whole dataset (i.e. 243 patients and 636 features), while for Algorithm 1 the mean and standard deviation over 10-folds are provided.

\begin{table}[]
	\centering \caption{Comparative run time study of the models.}
	\begin{tabular}{|c|c|c|}
		\hline
		\multicolumn{2}{|c|}{Task}                          & Time ($mean \pm SD$) \\ \hline
		\multirow{4}{*}{Featutre Extraction} & Radiomic     & $10.08\pm0.51$ (Sec.)          \\ \cline{2-3} 
		& Region       & $0.44\pm0.01$ (Sec.)         \\ \cline{2-3} 
		& Parcellation & $3.26\pm0.18$ (Sec.)         \\ \cline{2-3} 
		& Habitat      & $24.71\pm1.21$  (Sec.)        \\ \hline
		\multicolumn{2}{|c|}{Feature pre-processing}        & $5.12\pm0.21$ (Sec.)         \\ \hline
		\multicolumn{2}{|c|}{Algorithm 1}                   & $4.23\pm0.02$ (Sec.)         \\ \hline
	\end{tabular}
	\label{tab:run_time}
\end{table}

\begin{algorithm}[]
	\label{algo:1}
	\DontPrintSemicolon
	\SetAlgoLined
	\BlankLine
	\SetKwInOut{Input}{Input}\SetKwInOut{Output}{Output}
	\Input{Survival data $(\mathbf{X'}\in \mathbb{R}^{n\times p'}, \mathbf{T}\in\mathbb{R}^{n}, \mathbf{\delta}\in\{0,1\}^{n})$, number of folds ($k$)}
	\Output{Cross validation $C$-index score and the best number of features $\lambda_0$.}
	\BlankLine
	Split the dataset into $k$ (possibly) equal size folds.\;
	\For{s = 1 to $k$}
	{
		Use the fold $s$ as test set and remaining $k-1$ folds as training set.\;
		\For{$\lambda = 1, \dots, p'$}
		{
			Select top $\lambda$ features based on the ranking generated by fitting a CoxPH model on the training dataset.\;
			
			Fit the three survival analysis models \textit{viz.} Coxnet, SSVM, and RSF on the training set with the top $\lambda$ features. This constitutes the neuroimaging signature.\;
			
			Use the three fitted models to calculate the $C$-index
			on the test fold.\;
			
			Record the $C$-index value in three tables (for Coxnet, SSVM, and RSF) having $k$ columns and $p'$ rows.\;
		}
	}
	Compute the mean and standard deviation of the $C$-index values over all the $k$ folds for each of the $p'$ feature subsets corresponding to each of the three tables.\;
	
	Return the cardinality of the feature subset $\lambda_0$ (best number of features) having highest mean cross validation $C$-index value among the three survival  analysis models \textit{viz.} Coxnet, SSVM, and RSF.			
	
	\caption{$k$-fold cross-validation with grid search based univariate feature selection}
\end{algorithm}

\subsection{Results \label{sec:results}}
\begin{figure}[]
	\begin{center}
		\includegraphics[width=1.0\linewidth]{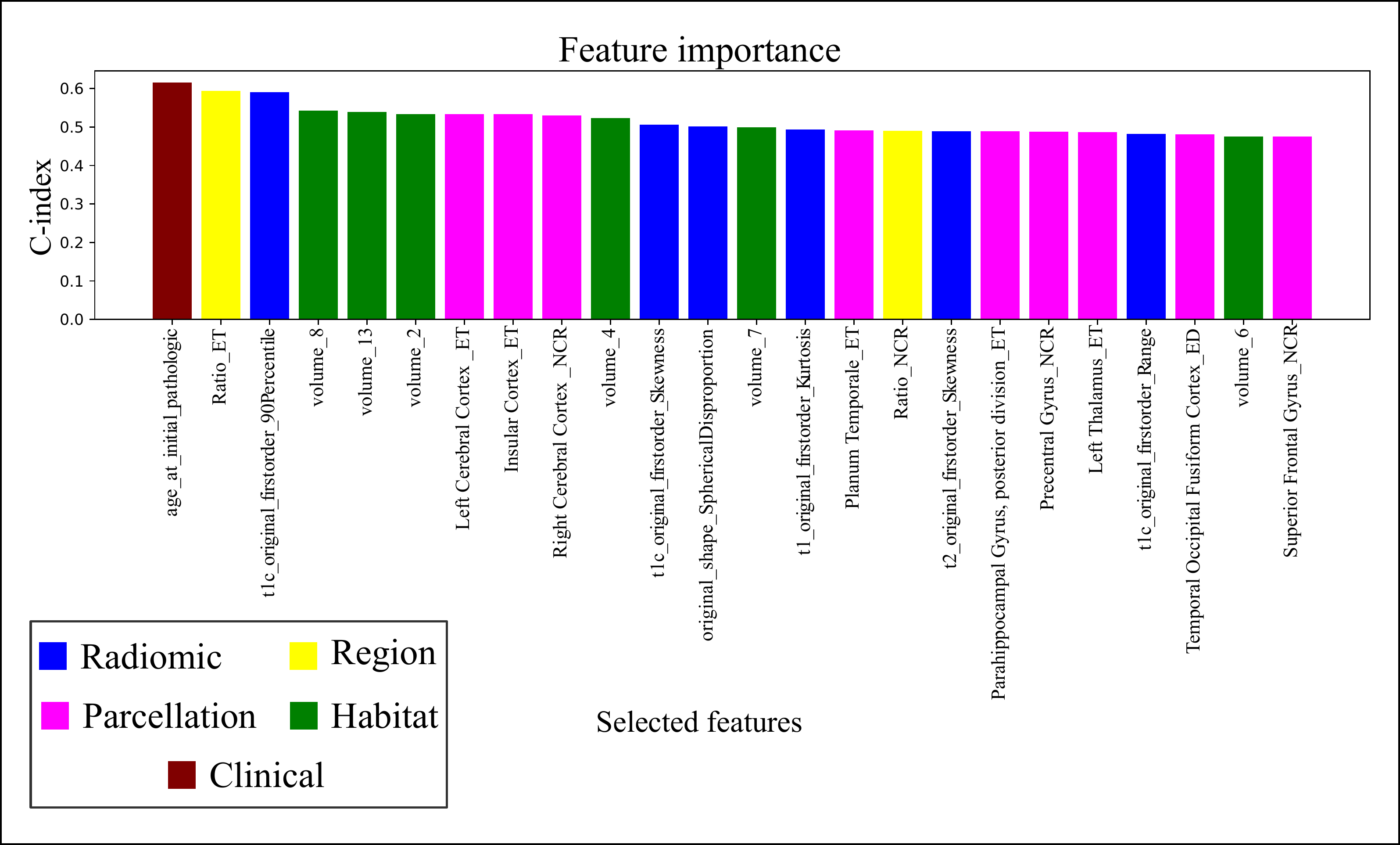}
	\end{center}
	\caption{Selected features by proposed algorithm.}
	\label{fig:feature_imp}
\end{figure}

Table \ref{tab:cv} reports the cross validation $C$-index scores, along with the cardinality of the best feature subset for the three survival analysis models \textit{viz.} Coxnet, SSVM, and RSF. It is observed from the table that RSF achieved the best mean cross validation $C$-index score ($0.82\pm10$) with $\lambda_0 = 24$. As these features may not be the same over all $k = 10$ folds, we computed their mean $C$-index individually over the $k$ folds to report the top performing features, and further study their clinical significance (as discussed in Sec. \ref{sec:clinical_sig}). As RSF obtained the best mean cross validation $C$-index score with $\lambda_0 = 24$ features, therefore the clinical importance of 24 features was studied \footnote{The 24 important features are computed with the objective of further studying of their clinical significance. As described in Step 5, we choose the top $\lambda$ features based on the training fold only, and measure their performance on the unseen test fold. Thereby, over different sets of training folds, the top $\lambda$ features selected by CoxPH may be different.}.  Thereby those features which performed consistently well, all over different folds, were identified (from the three feature families of Sec. \ref{section2}). Fig.~\ref{fig:feature_imp} plots the top $24$ features with respect to their cross-validation $C$-index value.

Among the $24$ selected features, age is found to be the most important. It is observed that $9$, $6$, $6$ and $2$ features are selected from the parcellation, habitat, radiomic and region-based groups, respectively. Moreover a combined feature set, including the most important features from each group, is seen to produce an optimal survival prediction model (last row of Table \ref{tab:cv}).

\begin{landscape}
\begin{table*}[]
	\centering \caption{Cross validation concordance index for survival prediction.}
	\begin{tabular}{ccccccc}
		\hline
		\multirow{3}{*}{\begin{tabular}[c]{@{}c@{}}Neuroimaging\\ features\end{tabular}} & \multicolumn{6}{c}{10-fold cross validation score}                                           \\ \cline{2-7} 
		& \multicolumn{2}{c}{Coxnet} & \multicolumn{2}{c}{SSVM}   & \multicolumn{2}{c}{RSF}    \\ \cline{2-7} 
		& $Mean\pm SD$ & $\# Features$ & $Mean\pm SD$ & $\# Features$ & $Mean\pm SD$ & $\# Features$ \\ \hline
		(C1) Radiomic                                                                        & $.70\pm.13$ & $28$         & $.69\pm.12$ & $29$         & $.70\pm.12$ & $28$         \\
		(C2) Region-based                                                                    & $.68\pm.12$ & $2$         & $.68\pm.12$ & $2$         & $.68\pm.10$ & $2$         \\
		(C3) Parcellation                                                                    & $.71\pm.10$ & $31$         & $.72\pm.12$ & $33$         & $.72\pm.11$ & $24$         \\
		(C4) Habitat                                                                         & $.71\pm.15$ & $28$         & $.73\pm.15$ & $25$         & $.73\pm.15$ & $24$         \\ \hline
		(C5) Radiomic+Region-based                                                           & $.73\pm.13$ & $27$         & $.73\pm.14$ & $25$         & $.74\pm.16$ & $25$         \\
		(C6) Radiomic+Parcellation                                                           & $.75\pm.12$ & $27$         & $.73\pm.13$ & $27$         & $.76\pm.12$ & $26$         \\
		(C7) Radiomic+Habitat                                                                & $.72\pm.12$ & $28$         & $.70\pm.13$ & $28$         & $.72\pm.13$ & $23$         \\
		(C8) Region-based+Parcellation                                                       & $.74\pm.10$ & $22$         & $.73\pm.12$ & $22$         & $.73\pm.10$ & $24$         \\
		(C9) Region-based+Habitat                                                            & $.74\pm.15$ & $29$         & $.73\pm.14$ & $29$         & $.74\pm.15$ & $29$         \\
		(C10) Parcellation+Habitat                                                            & $.74\pm.12$ & $28$         & $.74\pm.14$ & $28$         & $.75\pm.12$ & $28$         \\ \hline
		(C11) Radiomic+Region-based+Parcellation                                              & $.76\pm.13$ & $24$         & $.74\pm.11$ & $24$         & $.77\pm.11$ & $27$         \\
		(C12) Radiomic+Region-based+Habitat                                                   & $.76\pm.10$ & $25$         & $.75\pm.12$ & $25$         & $.77\pm.10$ & $27$         \\
		(C13) Radiomic+Parcellation+Habitat                                                   & $.78\pm.11$ & $27$         & $.77\pm.12$ & $25$         & $.80\pm.13$ & $27$         \\
		(C14) Region-based+Parcellation+Habitat                                               & $.76\pm.10$ & $26$         & $.75\pm.13$ & $28$         & $.78\pm.12$ & $25$         \\ \hline
		(C15) \textbf{Radiomic+Region-based+Parcellation+Habitat}                                      & $.79\pm.10$ & $22$         & $.78\pm.12$ & $22$         & $\mathbf{.82\pm.10}$ & $\mathbf{24}$ \\ \hline
	\end{tabular}
	\label{tab:cv}
\end{table*}
\end{landscape}

Testing was done to check whether the performance of the feature combination (C15), encompassing all the four feature categories, significantly improved the performance of the three survival analysis models. We compared all possible feature combinations consisting of two and three features groups and report the corresponding $p$-values in Table \ref{tab:t-test}. In case of feature group combinations of cardinality two, we observed a significant improvement (over those with individual feature sets C1-C4) in the performance for all the three survival analysis models. On comparing feature combination C13 (excluding region-based features) with the complete set of cardinality four C15, the improvement in performance for all the three survival models was found to be statistically not significant (ns). This implies that the ``region-based'' feature class is not that important for survival analysis. 

\begin{table*}[]
	\centering \caption{Statistical significance of feature combinations.}
	\begin{tabular}{|c|c|c|c|c|c|c|c|c|c|c|c|}
		\hline
		& C5               & C6               & C7               & C8               & C9               & C10              & \multirow{4}{*}{} & C11              & C12              & C13         & C14              \\ \cline{1-7} \cline{9-12} 
		Coxnet (C15) & \textless{}0.001 & \textless{}0.050 & \textless{}0.001 & \textless{}0.001 & \textless{}0.010 & \textless{}0.010 &                   & \textbf{ns}      & \textless{}0.050 & \textbf{ns} & \textless{}0.050 \\ \cline{1-7} \cline{9-12} 
		SSVM (C15)   & \textless{}0.010 & \textless{}0.010 & \textless{}0.001 & \textless{}0.010 & \textless{}0.010 & \textless{}0.050 &                   & \textless{}0.050 & \textbf{ns}      & \textbf{ns} & \textbf{ns}      \\ \cline{1-7} \cline{9-12} 
		RSF (C15)    & \textless{}0.001 & \textless{}0.001 & \textless{}0.001 & \textless{}0.010 & \textless{}0.001 & \textless{}0.001 &                   & \textless{}0.001 & \textless{}0.001 & \textbf{ns} & \textless{}0.050 \\ \hline
	\end{tabular}
	\label{tab:t-test}
\end{table*}

\subsection{Clinical significance of selected features \label{sec:clinical_sig}}
In this section we analyze the different groups of features for their utility towards survival analysis. The experiments suggest agreement with existing clinical findings, while also providing suggestions about possible wet lab explorations. 

\subsubsection{Radiomic and region} 
\begin{figure}[]
	\begin{center}
		\includegraphics[width=1.0\linewidth]{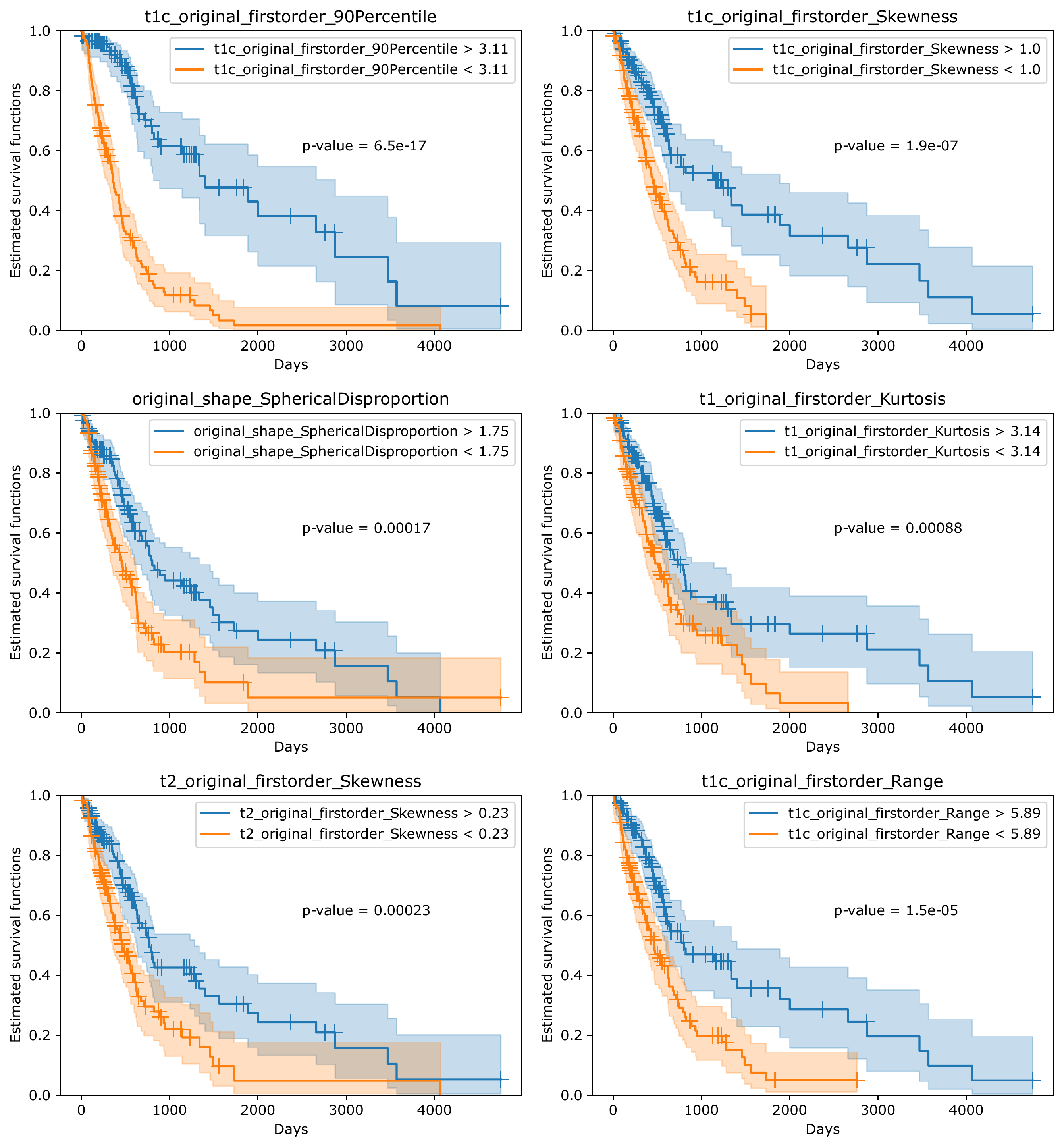}
	\end{center}
	\caption{Kaplan-Meier survival curves, comparing the survival rate of the $243$ glioma patients with respect to the $8$ radiomic features (with the spread of the colors representing the confidence intervals).}
	\label{fig:km_radiomic}
\end{figure}

Of the $24$ selected features of Fig. \ref{fig:feature_imp}, there were $6$ and $2$ features from the radiomic and region-based feature families. Among the $6$ radiomic features, $5$ belong to the GLS and $1$ to the GSS categories. Of the $5$ GLS features $3$, $1$, $1$ features were obtained from $T1C$, $T1$ and $T2$ sequences, respectively. In order to visualize the discriminative power of each selected feature, we divided the samples into two patient cohorts based on a threshold which was chosen as the corresponding median value of that feature (as in Ref. \cite{chaddad2018prediction}). Survival curves for each of these two groups [feature value $\ge$ median(feature), feature value $<$ median(feature)] were generated using the Kaplan-Meier estimator\footnote{Kaplan-Meier plots are used to visualize survival curves and the Log-rank test to compare the survival curves of two or more groups. The Log-rank test is the most widely used method of comparing two or more survival curves. The null hypothesis (H0) of the testing procedure is that there is no overall difference between the two (or more) survival curves. Under this H0, the log-rank statistic is approximately a chi-square.}. We used the log-rank test to determine whether there is a statistically significant difference between the patients having higher and lower values of the feature \cite{kleinbaum2012kaplan}. The Kaplan-Meier survival curves for the $6$ radiomic features are illustrated in Fig. \ref{fig:km_radiomic}. Features with $p$-value $< 0.05$  are considered to be statistically significant.

It is clearly evident from Fig. \ref{fig:km_radiomic}, that the selected radiomic features strongly correlate with the survivability of the patient. Note that most of the radiomic features are from the $T1C$ MR sequence, and $T1C$ quantifies the presence of enhancing tumor. This is typically found in HGGs \cite{blumenthal2017classification}, which exhibit poorer overall survival. 

\begin{figure}[]
	\begin{center}
		\includegraphics[width=1.0\linewidth]{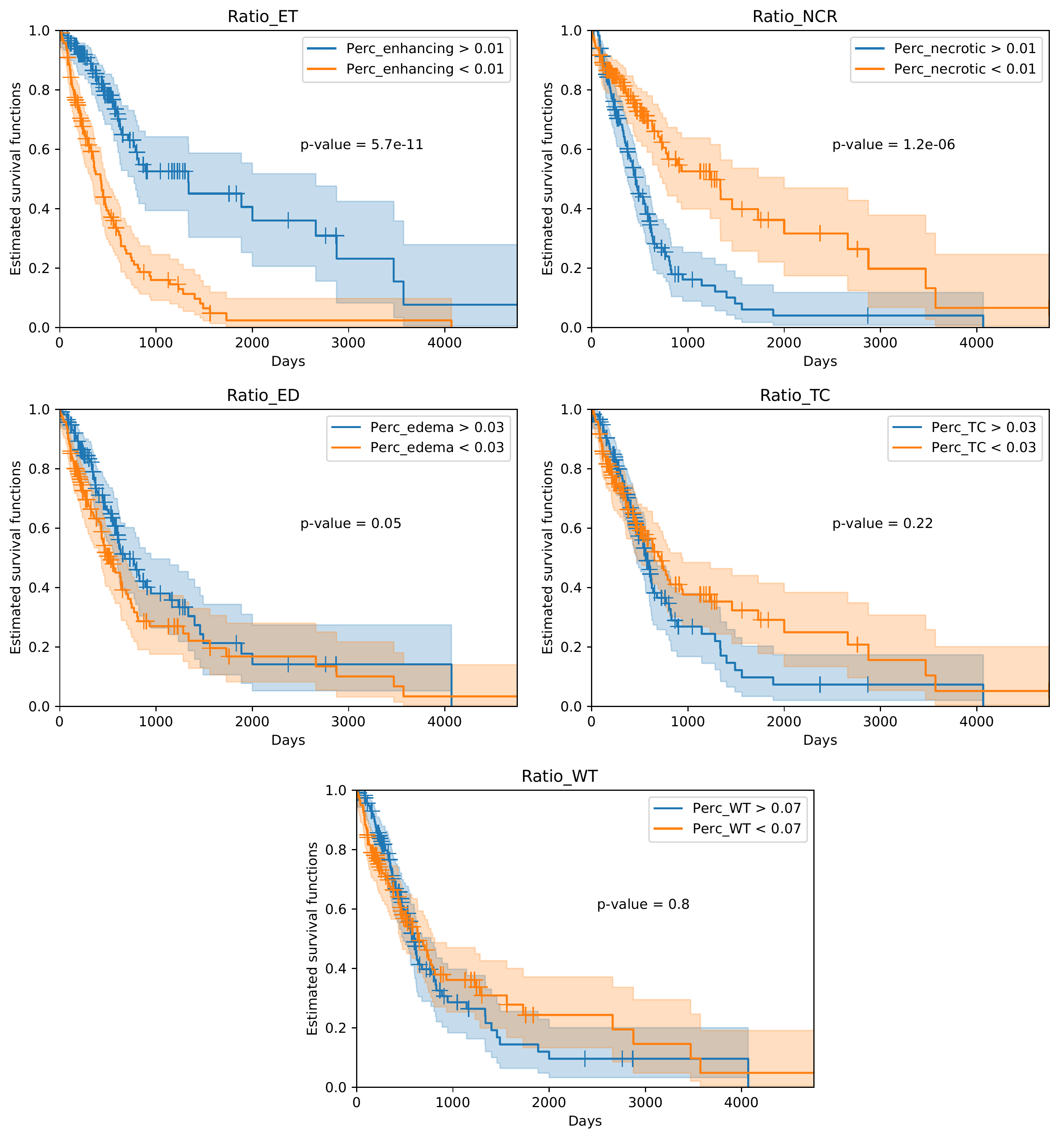}
	\end{center}
	\caption{Kaplan-Meier survival curves, comparing the survival rate of the $243$ glioma patients with respect to the $5$ tumor region-based features (with the spread of the colors representing the confidence intervals).}
	\label{fig:km_region}
\end{figure}

Among the five tumor volume ratios from the region features (``Ratio\_WT'', ``Ratio\_TC'', ``Ratio\_NCR'', ``Ratio\_ED'' and ``Ratio\_ET''), only two (``Ratio\_ET'', and ``Ratio\_NCR'') were selected by the CoxPH model. Again the Kaplan-Meier survival curves were generated for each of these five features, by dividing the patients into two groups based on the median value of the corresponding feature. It is obvious from Fig. \ref{fig:km_region} that only these two selected features effectively contribute towards differentiating survivability of the glioma patients.  

\subsubsection{Habitat} Six habitat volumes, viz. ``volume\_2'', ``volume\_4'', ``volume\_6'', ``volume\_7'', ``volume\_8'', and ``volume\_13'', were found to be most important for the survival model. Among these three habitat region volumes (``volume\_4'', ``volume\_7'' and ``volume\_13'') are corroborated to be significant for determining the overall survival in glioma patients \cite{dextraze2017spatial}. In Ref. \cite{blumenthal2017classification} the authors studied the importance of tumor habitats to predict survival in HGG patients only. On the other hand, we included LGG patients along with HGG; thereby, more habitat volumes were found to contribute towards differentiating survival prediction in both grades in our experiment. To study the correlation between the five habitat volumes with different tumor sub-regions, we used a pairwise Spearman rank test \cite{mukaka2012guide}. The corresponding correlation coefficient ($r_s$) and \textit{p}-value are provided in Table \ref{tab:habitat_corr}. As claimed in Ref. \cite{dextraze2017spatial}, we also observed a strong correlation between habitat 4 and necrosis ($r_s = 0.73$, $p < 0.00001$). Besides habitat 4 we found (i) habitats 2 \& 13 to also possess a strong positive correlation with NCR, (ii) habitats 8 \& 7 to have strong correlation with ET, and (iii) habitat 8 to be the only habitat having a strong correlation with both ET and ED. Habitats 2, 4, and 7 have strong positive correlation with TC and WT, whereas habitat 13 has a strong positive correlation with only TC.
\begin{landscape}
\begin{table*}[]
	\centering
	\caption{Pair-wise Spearman rank correlation between the six habitats and the three intra-tumoral components.}
	\begin{tabular}{|c|c|c|c|c|c|}
		\hline
		\diagbox[width=10em]{Habitat\\~}{Intra-tumoral\\components}
		& \begin{tabular}[c]{@{}c@{}}ET\\ $r_s$, $p$\end{tabular} & \begin{tabular}[c]{@{}c@{}}NCR\\ $r_s$, $p$\end{tabular} & \begin{tabular}[c]{@{}c@{}}ED\\ $r_s$, $p$\end{tabular} & \begin{tabular}[c]{@{}c@{}}TC\\ $r_s$, $p$\end{tabular} & \begin{tabular}[c]{@{}c@{}}WT\\ $r_s$, $p$\end{tabular} \\ \hline
		2       & $-0.06$, not significant                                              & $\mathbf{0.56}$, \textit{p} $< 0.00001$                                             & $0.40$, \textit{p} $< 0.00001$                                            & $\mathbf{0.56}$, \textit{p} $< 0.00001$                                            & $\mathbf{0.62}$, \textit{p} $< 0.00001$                                            \\ \hline
		7       & $\mathbf{0.51}$, \textit{p} $< 0.00001$                                          & $0.21$, \textit{p} $< 0.00100$                                              & $0.44$, \textit{p} $< 0.00001$                                            & $\mathbf{0.51}$, \textit{p} $< 0.00001$                                           & $\mathbf{0.57}$, \textit{p} $< 0.00001$                                            \\ \hline
		
		6       & $0.42$, \textit{p} $< 0.00001$                                            & $0.12$, not significant                                               & $0.43$, \textit{p} $< 0.00001$                                           & $0.40$, \textit{p} $< 0.00001$                                           & $0.48$, \textit{p} $< 0.00001$                                             \\ \hline
		4       & $0.20$, \textit{p} $< 0.00100$                                              &$\mathbf{0.73}$, \textit{p} $< 0.00001$                                             & ${0.44}$, \textit{p} $< 0.00001$                                            & $\mathbf{0.57}$, \textit{p} $< 0.00001$                                           & $\mathbf{0.73}$, \textit{p} $< 0.00001$                                          \\ \hline
		8       & $\mathbf{0.72}$, \textit{p} $< 0.00001$                                           & $-0.13$, not significant                                              & $\mathbf{0.53}$, \textit{p} $< 0.00001$                                            & $0.22$, 
		\textit{p} $< 0.00010$                                             & $0.43$, \textit{p} $< 0.00001$                                            \\ \hline
		13      & $-0.25$, not significant                                             & $\mathbf{0.58}$, \textit{p} $< 0.00001$                                            & $0.15$, \textit{p}  $< 0.05000$                                              & $\mathbf{0.55}$, \textit{p}  $< 0.00001$                                           & $0.47$, \textit{p} $< 0.00001$                                           \\ \hline
	\end{tabular}
	\label{tab:habitat_corr}
\end{table*}
\end{landscape}
\begin{figure}[!t]
	\begin{center}
		\includegraphics[width=1.0\linewidth]{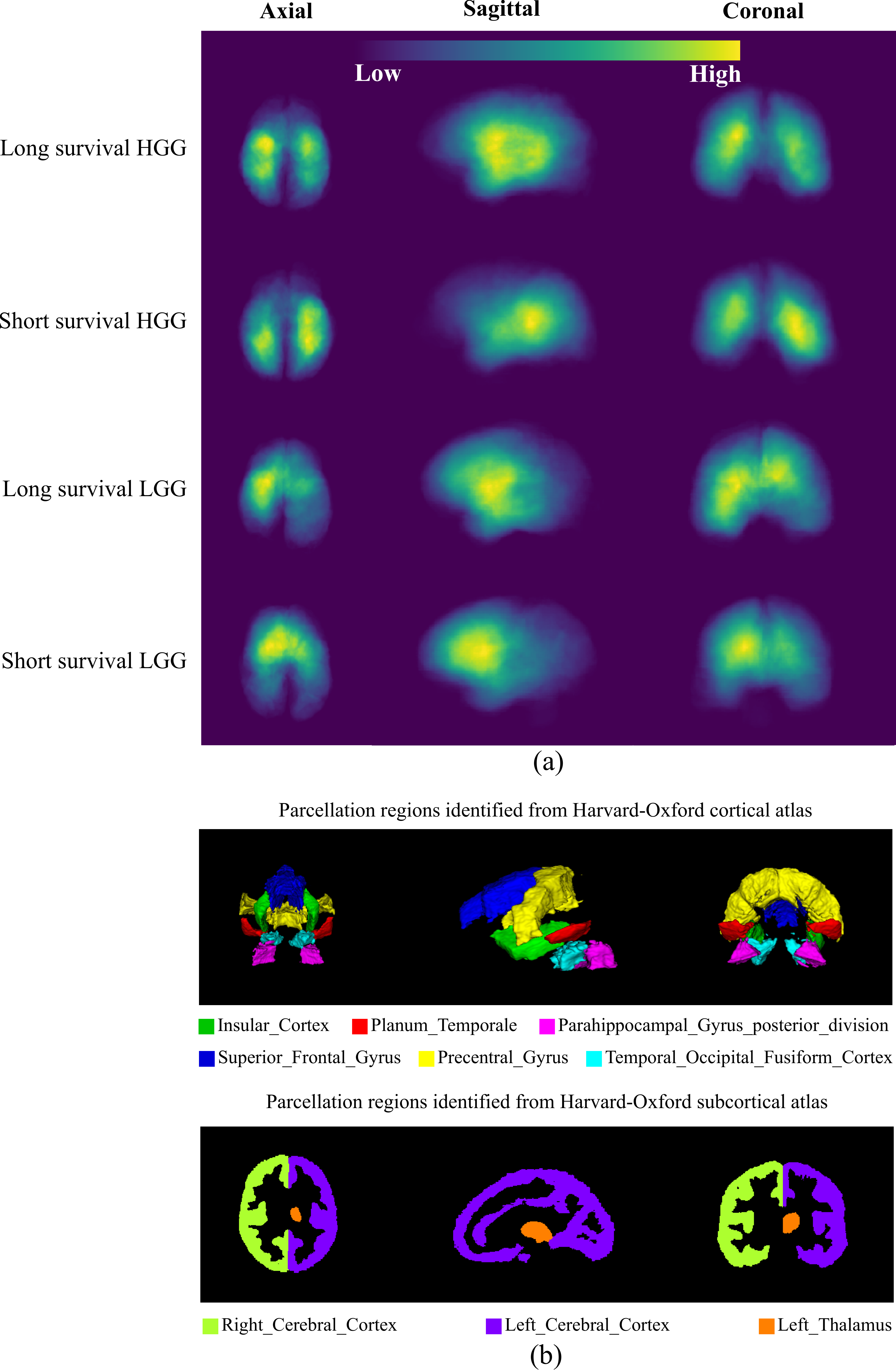}
	\end{center}
	\caption{(a) Heat map of the selected parcellation features, and the (b) corresponding important parcellation regions in the brain atlas.}
	\label{fig:parcellation_visual}
\end{figure}

\subsubsection{Parcellation} Among the nine parcellation atlas-based features, five (``Left Cerebral Cortex\_ET'', ``Insular Cortex\_ET'', ``Planum Temporal\_ET'', ``Parahippocampal Gyrus, posterior division\_ET'', and ``Left Thalamus\_ET'') indicate the percentage of ET volume present in the different brain parcellation regions. The remaining three features (``Right Cerebral Cortex\_NCR'', ``Precentral Gyrus\_NCR'' and ``Superior Frontal Gyrus\_NCR'') and one other feature (``Temporal Occipital Fusiform Cortex\_ED'') reflect the presence (in percentage) of NCR and ED volumes present in the different brain parcellation regions.

Heat maps were used to explore the interrelationship between the selected parcellation features and the patient survival, by computing the cumulative occurrence of the WT. This is illustrated in Fig. \ref{fig:parcellation_visual}. The HGG and LGG patients were divided into four groups  based on their corresponding median survival \textit{viz.} long survival HGG ($\ge$ median survival in HGG), short survival HGG ($<$ median survival in HGG), long survival LGG ($\ge$ median survival in LGG), short survival LGG ($<$ median survival in LGG). Visual inspection of the heatmaps helps associate the selected parcellation features with different survival groups.
 
It is observed that the features ``Insular Cortex\_ET'' and  ``Right Cerebral Cortex\_NCR'' can be associated with long survival in HGG. On the other hand, features ``Planum Temporal\_ET'', ``Left Cerebral Cortex\_ET'' and ``Left Thalamus\_ET'' are found to be visually correlated with short survival in HGG. While features ``Precentral Gyrus\_NCR'' and  ``Right Cerebral Cortex\_NCR'' can be associated with long survival in LGG patients, the ``Superior Frontal Gyrus\_NCR'' is observed to have a strong visual correlation with short survival in LGG.

\section{Conclusion \label{section5}}
An exhaustive study was undertaken to develop novel noninvasive imaging biomarkers, that can be useful for guiding adaptive therapeutic strategies for gliomas. Experiments demonstrated the effectiveness of these neuroimaging feature classes, along with the classical radiomic features, for survival analysis; based on a retrospective study using pre-operative multi-modal MRI data from multiple institutions. 

Best performance was achieved by combining all the four feature classes, \textit{viz.} Radiomic, Region-based, Habitat and Parcellation families. It was observed that a majority of the top ranking features belonged to the Habitat and Parcellation categories. Most of the selected radiomic features were from the GLS class. Interestingly the Region-based features, when considered separately, performed poorly; but in combination with the other three feature classes (\textit{viz.} Radiomic, Habitat and Parcellation) resulted in considerable improvement in terms of survival prediction. We note that when looking at different subsets of features chosen on all the data the accuracy may be optimistic.

The proposed neuroimaging biomarkers have the potential to be clinically useful, since they are interpretable and easy to compute. The availability of a large standard data set would allow us to fully assess their generalizability, and thereby improve the performance of these features based on the new unseen data set.


\bibliographystyle{unsrt} 
\bibliography{mybib}

\end{document}